\documentclass[%
 reprint,
superscriptaddress,
 amsmath,amssymb,
]{revtex4-2}

\usepackage{graphicx}
\usepackage{dcolumn}
\usepackage{bm}
\usepackage{soul}

\usepackage[counterclockwise, figuresleft]{rotating}
\usepackage{multirow}
\usepackage{longtable}
\usepackage{enumitem}

\begin{document}
\onecolumngrid
\title{Doppler orbital correction for the targeted search of known pulsars with the 5-vector method}

\author{Luca D'Onofrio} \email{ldonofrio@roma1.infn.it}
\affiliation{INFN, Sezione di Roma, I-00185 Roma, Italy} 
\author{Rosario De Rosa}
\affiliation{Università di Napoli “Federico II”, I-80126 Napoli, 
Italy}\affiliation{INFN, Sezione di Napoli, I-80126 Napoli, Italy} 
\author{Cristiano Palomba}
\affiliation{INFN, Sezione di Roma, I-00185 Roma, Italy}
\author{Paola Leaci}\affiliation{INFN, Sezione di Roma, I-00185 Roma, Italy}
\affiliation{Università di Roma "Sapienza", I-00185 Roma, Italy}


\begin{abstract}
Pulsars are promising targets for the first detection of continuous gravitational waves (CWs) by the LIGO-Virgo-KAGRA (LVK) Collaboration. Targeted searches leverage critical information derived from continued monitoring of pulsars related to the spin evolution, astrometry, and binary orbit. This information maximizes the probability to detect CW signals using full coherent methods. The 5-vector method is one of the three pipelines used by the LVK Collaboration for the targeted search of CWs from known pulsars. In this letter, we describe the first implementation and test of an heterodyne method to correct the orbital Doppler effect for pulsars in binary systems for the 5-vector targeted search.
\end{abstract}
\maketitle
\onecolumngrid
\section{Introduction}
Pulsars are rapidly rotating neutron stars primarily observed in the radio, X-ray, and $\gamma$-ray spectral bands. Continuous monitoring in the electromagnetic spectrum yields precise measurements of pulsar's celestial coordinates and rotational parameters. These details are vital for the search for continuous gravitational wave (CW) emissions by the LIGO-Virgo-KAGRA (LVK) detectors. Notably, non-axisymmetric mass distribution in spinning neutron stars can generate quasi-monochromatic gravitational wave radiation, with time scales on the order of months or years \cite{ornella}. In the simplest model, known as the "bumpy" neutron star model, the solid crust has a small deformation that entails the emission of a quasi-monochromatic signal with frequency at exactly twice the source's rotation frequency, $f_{gw}=2 f_{rot}$.

Understanding the source's parameters facilitates the application of full coherent methods, maximizing the sensitivity of CW analysis in targeted searches. The most recent targeted search by the LVK Collaboration in \cite{O3targ} analyzed 236 known pulsars considering O3  LIGO and Virgo datasets. This search yielded no evidence of a CW signal but set upper limits on the amplitude for each pulsar and, with an estimation of the distance, on the ellipticity.

The 5-vector method, proposed in \cite{2010} and improved in \cite{2014,mio2}, is one of three pipelines employed by the LVK Collaboration for the analysis of known pulsars. It utilizes a frequentist approach based on a matched filter in the frequency domain, which accounts for the Earth's sidereal modulation that splits the expected CW signal into five frequencies. Combined with the Band Sample Data (BSD) framework \cite{2019}, this method significantly reduces computational costs, allowing each pulsar's analysis to be completed within minutes on a common laptop. The BSD framework \cite{2019} consists of band-limited, down-sampled time series of the detector calibrated data, called BSD files, that covers 10$\,$Hz and spans 1 month of the original data. Using the BSD libraries, it is possible to extract frequency sub-bands or time sub-periods in a flexible way, reducing the computational cost of the analysis.

This letter presents the first implementation and testing of the heterodyne correction for the Doppler effect resulting from the orbital motion of pulsars in binary systems. The implementation written in MATLAB is included in the BSD library and generalize the Doppler and spin-down correction for isolated pulsar. This achievement enhances the capabilities of the 5-vector method, extending its applicability to binary systems.

\subsection{The 5-vector formalism}
In the 5-vector formalism, the expected CW signal $h(t)$ can be written as \cite{2010}:
\begin{equation}\label{compsign}
h(t)=H_0(H_+ A_+ + H_\times A_\times)e^{i\Phi(t)}
\end{equation}where\begin{equation}
H_+=\frac{\cos(2\psi)-j\eta \sin(2\psi)}{\sqrt{1+\eta^2}} \qquad \text{and} \qquad H_\times=\frac{\sin(2\psi)+j\eta \cos(2\psi)}{\sqrt{1+\eta^2}} \,,
\end{equation}
while $\eta$ and $\psi$ are the polarization parameters and $H_0$ the amplitude. The two functions $A_{+/\times}$ entail the detector response to the coming CW signal. 

$h(t)$ in equation \ref{compsign} is the product of two terms: the first, $(A_+H_+ + A_\times H_\times)$, is a “slow” amplitude and phase modulation at the Earth sidereal frequency while the exponential term is a “fast” term, due to the intrinsic source frequency and phase model.

The phase $\Phi(t)$ in \ref{compsign} shows a time dependence due to different phenomena that modulate in time the received signal frequency. The main physical phenomena to be considered are the spin-down and the Doppler effect due to the Earth motion and in case of binary systems, due to the source orbital motion. 
\subsection{The BSD heterodyne correction}\label{sec:et}
Different approaches \cite{velapulsar,eterodina} have been proposed to de-modulate the strain data in order to enhance the signal-to-noise ratio of the expected CW signal. In the 5-vector context, in \cite{res_bin} it is described a binary time-domain corrections to directed narrowband searches through the stroboscopic resampling. In this letter, we consider an heterodyne correction that needs precise knowledge of the source parameters, as in the case of targeted search.

In the BSD framework, detector's data are ready to be corrected without the need to apply filters or to re-sample, since this is already done by construction. This difference is useful in order to have a more general data framework which can be used also for other searches if needed. 

The signal phase shift due to the source spin-down can be written as:
\begin{equation}
\Phi_{sd}=2\pi \int_{\tau_{ref}}^t \left[\dot{f}(t'-\tau_{ref}) + \frac{1}{2}\ddot{f}_0(t'-\tau_{ref})^2+\ldots\right]dt' \,.
\end{equation}The corresponding phase factor for the Doppler correction is (up to a constant term):
\begin{equation}
\Phi_{d}=2\pi \int_{\tau_{ref}}^t f_0(t')\frac{\vec{r}\cdot \hat{n}}{c}dt'\approx \frac{2\pi}{c} p_{\hat{n}}(t)f(t)
\end{equation}where $p_{\hat{n}}(t)$ is the position of the detector in the chosen reference frame, projected along the source sky position $\hat{n}$ referred to the Solar System Barycenter (SSB).

The total signal phase correction can be written as the sum of the spin-down and the Doppler contributions:
\begin{equation}\label{phiref}
    \Phi_{corr}(t) =\Phi_{sd}(t) +\Phi_{d}(t)
\end{equation}
Heterodyne de-modulation is then applied by multiplying the data by the exponential factor $e^{-i\Phi_{corr}(t)}$.
This technique is useful for correcting pulsar signals for the phase modulation caused by the spin-down/Doppler shift and hence to  precisely  unwind  the apparent  phase  evolution  of the source.

Residual modulations might also be present due to inappropriate modeling of the frequency evolution (higher order spin-down terms, source frequency glitches) or to parameter uncertainties like a not perfect estimation of source position parameters.  

\section{Binary system correction}\label{sec:etbin}
If the source is in a binary system, the phase correction  in \ref{phiref} should also include the term $\Phi_{bin}(t)$ that takes into account the shift in the arrival time of the signal due to the motion of the source within the binary system. Hence, there is an additional phase correction due to the orbital Doppler effect equal to the Roemer delay caused by the binary motion. It can be written as:
\begin{equation}
    \Phi_{bin} (t)=-\frac{R(t)}{c}
\end{equation}where $R$ is is the radial distance of the CW emitting source from the binary barycenter (BB) with $R>0$ if the source is further away from us than the BB. Following \cite{deltabin}, 
\begin{equation}
    \frac{R(t)}{c}=a_p\left[ \sin \omega\left( \cos E -e \right) + \cos \omega \sin (E\sqrt{1-e^2}) \right]
\end{equation}where $a_p=a \sin \iota/c$ is the projected semi-major axis ( with $\iota$ the inclination angle and $a$ the semi-major axis) and $e$ the orbital eccentricity. The eccentric anomaly $E$ is defined by the Kepler's equation:
\begin{equation}
    t -t_p\approx \frac{P}{2\pi}\left( E-e\sin E \right)
\end{equation}with $P$ the orbital period and $t_p$ the time of periapse. In these relations, we have considered the detector in the SSB since knowing the source sky position, we can rewrite the timing relation (see \cite{deltabin} for more details). 


In the case of low eccentric orbit ($e\to 0$), the phase model can be simplified. Considering $E(t)=E_0(t)+eE_1(t)+...$ and the Kepler's equation, the Roemer delay to leading order in $e$ is:
\begin{equation}
    \frac{R(t)}{c}=a_p \left[ \sin \psi(t) + \frac{k}{2} \sin 2\psi(t) -\frac{\beta}{2}\cos 2\psi(t) -\frac{3}{2}\beta \right]
\end{equation}with
\begin{equation}
    k=e\cos\omega  \qquad \text{and} \qquad   \beta=e\sin \omega
\end{equation}
and the mean orbital phase
\begin{equation}
    \psi(t)=\Omega (t-t_{asc})
\end{equation}measured from the time of ascending nodes $t_{asc}$. For small eccentricity and mean orbital angular velocity $\Omega=\frac{2\pi}{P}$, $t_{asc}$ is:
\begin{equation}
    t_{asc}=t_p-\frac{\omega}{\Omega}
\end{equation}

\section{Test with hardware injections}\label{sec:HI}
Starting from this phase model, we use the heterodyne method to remove the signal modulation due to the orbital motion of the source and to apply the 5-vector pipeline to binary systems. Indeed, after the Doppler and spin-down corrections, there is a residual modulation in amplitude and phase due to the Earth sidereal modulation. The residual modulation is in the functions $A_{+/\times}$ that splits the signal in the 5 frequencies $f_{\text{gw}},f_{\text{gw}} \pm \Omega_\oplus,f_{\text{gw}} \pm 2 \Omega_\oplus $ where $\Omega_\oplus$ is the Earth's sidereal angular frequency.  The data 5-vector $\textbf{X}$ and the signal template 5-vectors  $\textbf{A}^{+/\times}$ are defined as the Fourier transforms of the data and of the template functions $A_{+/\times}$ at the 5 frequencies where the signal power is split.
The 5-vector method defines two matched filters between the data $\textbf{X}$ and the signal templates $\textbf{A}^{+/\times}$ vectors, used in order to maximize the signal-to-noise ratio:
\begin{equation}\label{estim}
\hat{H}_+=\frac{\textbf{X}\cdot \textbf{A}^+}{|\textbf{A}^+|^2} \qquad \text{and} \qquad \hat{H}_\times=\frac{\textbf{X}\cdot \textbf{A}^\times}{|\textbf{A}^\times|^2}
\end{equation} These two matched filters are used \cite{2010} to estimate the signal parameters.

The implementation of the orbital Doppler shift considers two different cases according to the binary pulsars model provided by the astronomers. Indeed, there are two main models for the orbital parameters \cite{tempo23}: the T2 model that returns the parameters ($P$, $a_p$, $t_p$, $\omega$, $e$) and the ELL1 model that returns the parameters ($P$, $a_p$, $t_{asc}$, $k$, $\beta$). In the low-eccentricity limit,  the T2 model returns strongly correlated values for $t_p$ and $\omega$, and the ELL1 model was developed for pulsars in such orbits.

To test the implementation of this correction, we consider the hardware injections present in O3 data.
Hardware injections are simulated signals added to the detectors data physically displacing the detectors’ test masses \cite{HI}. Differential displacement of the test masses mimics the detectors’ response to a GW signal. 

During the O3 run, $17$ hardware injections that simulate different CW signals from spinning neutron stars were added in the two LIGO detectors. The list of injected pulsar parameters can be found in \cite{HIlist}. For Virgo, no CW injections were performed during O3 (there were some injections during O3b but the signal was removed a posteriori in the strain data).

In this Section, we consider two different hardware injections to test the heterodyne correction. We report the results for the injected pulsar P03 (SNR(1 yr) $\approx$ 30) that simulates an isolated spinning neutron stars with $f_{gw}\simeq 108.8$ Hz and for P16 (SNR(1 yr) $\approx$ 68), a neutron star in binary system with $f_{gw}\simeq 234.5$ Hz.
\begin{figure}
\centering
\includegraphics[scale=0.6]{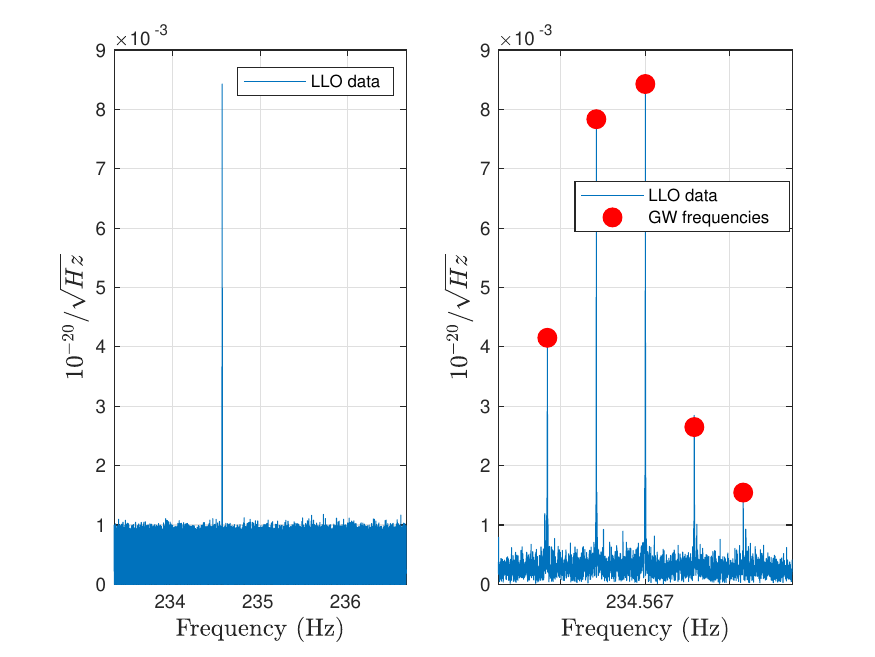}
\caption{\small{Amplitude spectral density for P16 and LHO O3 data after the corrections for Doppler and spin-down effect. The right plot is the zoom around the GW frequency; the red dots show the expected 5 frequencies due to the Earth sidereal motion. Central peak frequency is the expected GW frequency.}}
\label{fig:HI16ASDH}
\end{figure}
Figures \ref{fig:HI16ASDH} shows the amplitude spectral density in the frequency band around the expected GW frequency for P16 and for the Hanford detector after the Doppler and spin-down corrections using the BSD heterodyne method. The red dots are the theoretical expected five peaks in the frequency domain for the CW signal due to the sidereal modulation. It is important to note that the formalism used to construct the hardware injections' signal is independent from the 5-vector method. 

In Table \ref{tab:result}, there are the results for the parameters estimation (see \cite{2010}) obtained for the two hardware injections considering the O3 datasets of the two LIGO detectors. 
\begin{table}[t]
\setlength{\arrayrulewidth}{0.2mm}
\setlength{\tabcolsep}{12pt}
\renewcommand{\arraystretch}{3}
\renewcommand{\arraystretch}{1.5}
  \begin{center}
    	\begin{tabular}{|c| c| c| c| c|}
    	\hline
\textbf{Hardware Injection} & Detector & $\epsilon_{H_0}$     & $\epsilon_\eta$      & $\epsilon_\psi$ \\
    	\hline
\multirow{2}{*}{P03}        & LLO   & 0.93  & 0.54\% &  1.49\%     \\
                            & LHO   & 0.95 & 1.27\%  & 0.39\%      \\
    	\hline
\multirow{2}{*}{P16}        & LLO   & 0.91  & 0.66\% &  -     \\
                            & LHO   & 0.93 & 0.90\% & -     \\
    	\hline                            
\end{tabular}
\end{center}
  \caption{ Table of the parameters mismatch for the hardware injections P03 and P16 analyzing O3 data from the two LIGO detectors.}
  \label{tab:result}
\end{table}
\\The parameter mismatch definition depends on the considered parameter: $\epsilon_{H_0}$ is the ratio between the estimated $\hat{H}_0$ and injected amplitude $H_0$; $\epsilon_{\eta}$ is the normalized relative errors, defined as $\epsilon_{\eta}=(\eta-\hat{\eta})/2$, "normalized" in such a way that the maximum relative error between two values is $1$ ( $\eta_{max} = +1$ and $\eta_{min} = -1$); while $\epsilon_{\psi}=(\psi-\hat{\psi})/\pi$ since $\psi_{max}=\pi/2$ and $\psi_{min}=-\pi/2$.

The small discrepancies (below 10\%) in Table \ref{tab:result} for the amplitude  fall within the uncertainties of the actuation system used for the injections. For P16, the $\psi$ parameter is not well defined, since $\eta \approx 1$ and the signal is circularly polarized.

\section{Conclusion}
Pulsars are promising targets for the first CWs detection by the LVK Collaboration since targeted searches maximize the sensitivity. The limits of these searches rely in the dependence on the emission model. Indeed, different emission scenarios entail a different proportionality between the CW frequency and the source rotation frequency. Narrow-band searches can be performed to account for a different and unknown relation between the CW frequency and the source rotation frequency.

This letter shows the first implementation and test of the orbital Doppler correction for pulsars in binary systems in the 5-vector targeted search using an heterodyne method. The code in MATLAB considers the phase shift due to the orbital Doppler effect, described for example in \cite{deltabin}, and it is included in the BSD library. Using this implementation, the 5-vector pipeline can now be applied for the targeted search of pulsars in binary systems.

In the near future, we must conduct a detailed analysis of the potential impacts of large uncertainties in the binary orbital parameters on the sensitivity of targeted search. Our objective is to quantify the phase deviation resulting from the interplay and correlation of these uncertainties and their subsequent impact on the signal-to-noise ratio. We aim to establish a threshold for the phase deviation as a function of the parameters uncertainties, which will serve as a limiting factor for the sensitivity loss.

\bibliographystyle{unsrtnat}
\bibliography{biblio}

\end{document}